\documentclass[sigconf]{acmart}

\AtBeginDocument{%
  \providecommand\BibTeX{{%
    \normalfont B\kern-0.5em{\scshape i\kern-0.25em b}\kern-0.8em\TeX}}}

\copyrightyear{2023}
\acmYear{2023}
\setcopyright{rightsretained}
\acmConference[SIGIR '23]{Proceedings of the 46th International ACM SIGIR Conference on Research and Development in Information Retrieval}{July 23--27, 2023}{Taipei, Taiwan}
\acmBooktitle{Proceedings of the 46th International ACM SIGIR Conference on Research and Development in Information Retrieval (SIGIR '23), July 23--27, 2023, Taipei, Taiwan}\acmDOI{10.1145/3539618.3591985}
\acmISBN{978-1-4503-9408-6/23/07}

\usepackage[utf8]{inputenc}
\usepackage{amsthm}
\usepackage{multicol}

\usepackage{booktabs,enumitem,multirow}
\usepackage[normalem]{ulem}
\usepackage{framed,fancybox,ascmac}
\usepackage{wrapfig}
\usepackage{enumitem}
\usepackage{diagbox}

\usepackage{mathtools}
\usepackage{apxproof}

\usepackage{comment}
\usepackage{url}
\usepackage[caption=false]{subfig}

\usepackage[colorinlistoftodos,prependcaption]{todonotes}
\usepackage{xspace}

\usepackage{bm}

\usepackage[nameinlink]{cleveref}
\usepackage{nameref}

\usepackage{tcolorbox}

\usepackage{lineno}

\newtheorem{theorem}{Theorem}[section]
\newtheorem*{theorem*}{Theorem}

\theoremstyle{definition}
\newtheorem{definition}[theorem]{Definition}

\crefname{theorem}{Theorem}{Theorems}
\crefname{lemma}{Lemma}{Lemmas}
\crefname{claim}{Claim}{Claims}
\crefname{remark}{Remark}{Remarks}
\crefname{observation}{Observation}{Observations}
\crefname{corollary}{Corollary}{Corollaries}
\crefname{appendix}{Appendix}{Appendices}
\crefname{section}{Section}{Sections}
\crefname{algorithm}{Algorithm}{Algorithms}
\crefname{equation}{Eq.}{Eqs.}
\crefname{figure}{Figure}{Figures}
\crefname{table}{Table}{Tables}

\usepackage{algorithmicx}
\usepackage[ruled]{algorithm}
\usepackage[noend]{algpseudocode}
\algnewcommand\And{\; \textbf{and} \;}
\algnewcommand\Or{\; \textbf{or} \;}
\algnewcommand\To{\; \textbf{to} \;}
\algnewcommand\Continue{\textbf{continue}}
\algnewcommand\Not{\textbf{not}}

\algnewcommand{\algalign}[1]{\parbox[t]{\dimexpr\linewidth-\algorithmicindent}{#1\strut}}
\algrenewcommand\textproc{\textsl}

\DeclareMathOperator*{\argmax}{argmax}

\newcommand{\calD}{\mathcal{D}}

\newcommand{\bbN}{\mathbb{N}}

\newcommand{\bbZ}{\mathbb{Z}}

\newcommand{\bfP}{\mathbf{P}}

\newcommand{\E}{\mathbb{E}}

\newcommand{\sigmaldrt}{\sigmaldr_t}
\newcommand{\sigmaldr}{\bsigma^{\mathrm{LDR}}}
\newcommand{\sigmat}{\bsigma_t}
\newcommand{\bsigma}{\boldsymbol{\sigma}}
\newcommand{\one}{\mathbf{1}}
\newcommand{\doublebar}[1]{\bar{\bar{#1}}}
\newcommand{\doublebars}{\doublebar{s}}

\begin{document}

\title[SafeOLTR]{Exploration of Unranked Items \\ in Safe Online Learning to Re-Rank}

\author{Hiroaki Shiino}
\affiliation{
    \institution{CyberAgent, Inc.}
    \country{Japan}
}
\email{shiino_hiroaki@cyberagent.co.jp}

\author{Kaito Ariu}
\affiliation{
    \institution{CyberAgent, Inc.}
    \country{Japan}
}
\email{kaitoariu@gmail.com}

\author{Kenshi Abe}
\affiliation{
    \institution{CyberAgent, Inc.}
    \country{Japan}
}
\email{abe_kenshi@cyberagent.co.jp}

\author{Riku Togashi}
\affiliation{
    \institution{CyberAgent, Inc.}
    \country{Japan}
}
\email{rtogashi@acm.org}

\renewcommand{\shortauthors}{H. Shiino et al.}

\begin{abstract}
Bandit algorithms for online learning to rank (OLTR) problems often aim to maximize long-term revenue by utilizing user feedback.
From a practical point of view, however, such algorithms have a high risk of hurting user experience due to their aggressive exploration. 
Thus, there has been a rising demand for \emph{safe} exploration in recent years.
One approach to safe exploration is to gradually enhance the quality of an original ranking that is already guaranteed acceptable quality.
In this paper, we propose a safe OLTR algorithm that efficiently exchanges one of the items in the current ranking with an item outside the ranking (i.e., an unranked item) to perform exploration.
We select an unranked item \emph{optimistically} to explore based on Kullback-Leibler upper confidence bounds (KL-UCB) and safely re-rank the items including the selected one.
Through experiments, we demonstrate that the proposed algorithm improves long-term regret from baselines without any safety violation.
\end{abstract}

\ccsdesc[500]{Information systems~Learning to rank}
\ccsdesc[500]{Computing methodologies~Online learning settings}
\ccsdesc[500]{Computing methodologies~Learning from implicit feedback}

\keywords{Safety; online learning to rank; implicit feedback}


\maketitle
\section{Introduction}
\label{sec:intro}

Learning-to-rank (LTR) methods play a key role in delivering attractive content to
users living in the era of information overload, wherein new content is rushing into a database every day.
Systems must provide such new content quickly and accurately to users, who will be interested in it.
However, we often face a lack of information about new content, and therefore, exploration is essential, although it inevitably implies somehow baseless prediction to initially collect information, which can damage user satisfaction.
Thus, systems are in a dilemma---exploration is necessary, yet its \emph{safety} is indispensable.

Online learning-to-rank (OLTR)~\cite{yue2009interactively,hofmann2013balancing} is a promising approach to combat the information lack of new items by immediately reflecting fresh user feedback collected through a prediction-observation loop.
Several conventional studies have developed OLTR methods using item features~\cite{li2019online},
and this approach is also effective for handling new items.
On the other hand, features of new items may be unreliable in practice, particularly when there is an unexpected craze for a new one;
in this situation, features that were previously effective will not perform well.
Recent OLTR studies have explored methods based on ranking bandits~\cite{kveton2022value,combes2015learning,kveton2015cascading,kveton2015combinatorial,lagree2016PBMmultiple,komiyama2017position, magureanu2017onlline,zong2016cascading}.
Whereas this approach enables learning to rank new items without relying on item features, a click model is often assumed to learn item attractiveness from biased click feedback.
However, accurately specifying the click model behind user feedback is generally challenging, and hence such a model-specific approach may be \emph{unsafe} because misspecified models lead to inaccurate prediction and can hurt user satisfaction.
Recent studies have proposed click-model-agnostic algorithms~\cite{zoghi2017batchrank, lattimore2018toprank,li2020bubblerank,gauthier2022unirank}, which is safe in terms of model misspecification.
\citet{li2020bubblerank} proposed a click-model-agnostic method, \emph{BubbleRank}, which explicitly considers the safety of exploration in an OLTR setting,
where algorithms can leverage an original ranking generated by a method previously deployed in the production system.
Based on the definition of \citet{li2020bubblerank}, BubbleRank is ``safe'' in the sense that the ranking shown to a user does not substantially underperform an original ranking with high probability.  
Nevertheless, BubbleRank is basically designed for re-ranking of originally ranked items and cannot efficiently handle unranked items, which do not appear in the original ranking;
although the extension with the random exploration of unranked items is discussed, their na{\" i}ve strategy is statistically inefficient as shown in this paper.

In this paper, we develop an OLTR algorithm that can safely explore unranked items by extending BubbleRank~\cite{li2020bubblerank}.
To achieve safe exploration for unranked items without any preliminary information,
we utilize the Kullback-Leibler upper confidence bounds \cite{cappe2013kullback} (KL-UCB) as the optimistic confidence measure of item attractiveness.
To examine the effectiveness of our proposed algorithm in various scenarios, we conduct semi-simulate experiments on the real-world dataset.

\section{Related Work}
\label{sec:related}


Conventional OLTR algorithms can be classified into the click-model-specific approach~\cite{kveton2022value,combes2015learning,lagree2016PBMmultiple,komiyama2017position,kveton2015cascading,kveton2015combinatorial,magureanu2017onlline,zong2016cascading} and click-model-agnostic counterpart~\cite{zoghi2017batchrank, lattimore2018toprank, li2020bubblerank, gauthier2022unirank}. 
Model-specific algorithms assume a certain click model behind user feedback data to efficiently learn optimal rankings when users follow the assumed click model, e.g., position-based model (PBM)~\cite{lagree2016PBMmultiple,komiyama2017position} and the cascade model (CM)~\cite{kveton2015cascading,kveton2015combinatorial,zong2016cascading, craswell2008experimental}.
However, it can be \emph{unsafe} in the sense that
their theoretical guarantees do not hold when the assumed click model does not fit the actual user behavior~\cite{li2020bubblerank, gauthier2022unirank}.
By contrast, the model-agnostic counterpart only requires weak assumptions.
UniRank~\cite{gauthier2022unirank} is the state-of-the-art model-agnostic algorithm with excellent performance in terms of regret, but it does not consider safety constraints.
To achieve safe re-ranking without assuming click models,
\citet{li2020bubblerank} proposed BubbleRank, which has a severe limitation in handling unranked items because their proposed random exploration does not consider statistical efficiency.

The notion of safety in OLTR is related to that in conservative bandit algorithms.
In conservative bandit, the notion of safety is defined as a constraint on cumulative rewards~\cite{wu2016conservative, kazerouni2017conservative}; notably, at each round, algorithms are allowed to select arms that can cause high regret, as long as the constraint is respected throughout the entire rounds.
Beyond such a ``coarse-grained'' definition, some recent studies consider \emph{stage-wise} safety, which requires algorithms to be conservative in every round~\cite{moradipari2020stage, khezeli2020safe}.
This stage-wise definition is rather related to the safety of interest in this study.
Unfortunately, the existing algorithms are designed for linear bandit settings~\cite{khezeli2020safe, moradipari2020stage} and thus cannot be applied to the OLTR settings efficiently.

In this paper, we propose a model-agnostic algorithm 
inspired by BubbleRank and UniRank, which enables stage-wise safe re-ranking and exploration under unranked items.






\section{Problem Formulation}
\label{sec:problem_formulation}



An instance of a \textit{stochastic click bandit} is a tuple $(K,L,P_{\alpha},P_{\chi})$, where $L \in \bbN$ is the size of the set of all items $\calD$, $P_{\alpha}$ is a distribution over binary attraction vector $\{0,1\}^L$, and $P_{\chi}$ is a distribution over binary examination matrices $\{0,1\}^{\Pi_K(\calD)\times K}$, with $\Pi_K(\calD)$ is the set of all permutations of $K(\leq L)$ items from $\calD$. 

For $n\in \bbZ_{+}$, which is the set of non-negative integers, let $[n] := \{1,\ldots, n\}$.
At each round $t\in[T]$, an algorithm shows a ranking $\bsigma_t\in\Pi_K(\calD)$ to a user and observes the user's clicks $\{c_t(k)\}_{k=1}^{K} \in \{0,1\}^K$ on all positions in $\bsigma_t$.
Note that $\bsigma_t$ depends on the past history up to round $t-1$.
A position is clicked if and only if it is examined and the item at that position is attractive, that is, for any $k\in[K]$,
$c_t(k) = X_t(\bsigma_t,k)A_t(\sigmat(k))$,
where $X_t(\bsigma,k)\in \{0,1\}$ is the examination indicator of position $k$ in a ranking $\bsigma\in\Pi_K(\calD)$ at round $t$; $A_t(\bsigma(k))\in \{0,1\}$ is the attraction indicator of an item $\bsigma(k)$ at position $k$ in a ranking $\bsigma\in\Pi_K(\calD)$ at round $t$.
For notational simplicity, for $k>K$, we let $X_t(\bsigma, k) = 0$ and $c_t(k) = 0$.
Both $\{X_t(\bsigma,k):\bsigma\in\Pi_K(\calD),k\in[K]\}$ and $\{A_t(i):i\in[L]\}$ are stochastic and drawn i.i.d. from $P_{\chi}\bigotimes P_{\alpha}$.

The expected reward for an algorithm at round $t$ conditioned on a ranking $\bsigma$ is the summation of expected clicks: 
$$
r(\bsigma, \alpha, \chi(\bsigma)):= \sum_{k=1}^{L}\E[c_t(k)|\bsigma_t=\bsigma]=\sum_{k=1}^{L}\chi(\bsigma,k)\alpha(\bsigma(k)),
$$
where $\chi(\bsigma,k)=\E[X_t(\bsigma_t,k)|\sigmat = \bsigma]$ is the examination probability of position $k$ in $\bsigma$, $\alpha(i)=\E[A_t(i)]$ is the attraction probability of item $i$,
 $\chi(\bsigma)=\{\chi(\bsigma,k)\}_{k=1}^{K}$, and  $\alpha = \{\alpha(i)\}_{i=1}^{L}$.
The goal of this problem is then to minimize the \textit{cumulative expected regret}
\begin{align}
\label{eq:regret}
R(T) = T\cdot r^* - \E\left[\sum_{t=1}^{T}r(\sigmat,\alpha,\chi(\sigmat))\right],
\end{align}
where $r^*:=\max_{\bsigma\in\Pi_K(\calD)} r(\bsigma,\alpha,\chi(\bsigma))$ is the highest expected reward and the expectation is taken with respect to the rankings from the algorithm and the clicks.

Our problem requires the assumptions introduced by~\citet{li2020bubblerank}.
The assumptions hold in the CM and they also do in the PBM when the examination probability decreases with the position. 
Then the optimal ranking $\bsigma^*:= \argmax_{\bsigma\in\Pi_K(\calD)} r(\bsigma,\alpha,\chi(\bsigma))$ includes the item which has $k$-th highest attractiveness at position $k$.

\section{Proposed Method}
\label{sec:proposed}

Our algorithm, called KL-UCB-BR, is described in Algorithm~\ref{alg:fullbubblerank}.
KL-UCB-BR holds the following three rankings in each round $t$: a \emph{leader} ranking $\sigmaldrt$, a temporary ranking $\bsigma'_t$, and a display ranking $\bsigma_t$.
The leader ranking $\sigmaldrt$ is the interim best ranking estimated at round $t$; we initialize it to the original ranking $\bsigma_0$ if $t=1$ and, if $t>1$, to the top-$K$ partial ranking of the previous temporary ranking $\bsigma'_{t-1}([K])$.
The temporary ranking $\bsigma'_t$ is used to exchange items to safely reorder the item pairs; we initialize it with the current leader ranking $\sigmaldrt$, set a single unranked item to explore at the $(K+1)$-th position, and exchange items at consecutive positions to ensure the correct order.
The display ranking $\bsigma_t$ is presented to users, and user clicks can be observed on the items in it; we initialize it to the current temporary ranking $\bsigma'_t$ and update it by randomly exchanging items at consecutive positions if KL-UCB-BR is not confident about the order in their attractiveness.

As the criteria to safely reorder the item pairs to be correct, we utilize the following statistics for the exchanging items in the temporary ranking~\cite{li2020bubblerank}:
\begin{align*}
s_t(i,j) &:= \sum_{s=1}^{t-1}O_s(i,j)(c_s(\bsigma_s^{-1}(i))-c_s(\bsigma_s^{-1}(j))),\\
n_t(i,j) &:= \sum_{s=1}^{t-1}O_s(i,j),
\end{align*}
where $\bsigma^{-1}(i)$ is a position of item $i$ in a ranking $\bsigma$ and $O_s(i,j):=\one\left\{(i,j)\in \bfP_s(\bsigma_s)\right\}\one\{c_s(\bsigma_s^{-1}(i))\neq c_s(\bsigma_s^{-1}(j))\}$, with $\bfP_s(\bsigma)$ is the set of item pairs at odd-even/even-odd consecutive positions in a ranking $\bsigma$ in even/odd round $s$.
From Lemma~9 of~\citet{li2020bubblerank}, when $s_t(i,j)>\sqrt{n_t(i,j)\log(1/\delta)}$ with sufficiently small $\delta$, item $i$ is superior to item $j$ with high probability.

We select a single unranked item to explore in round $t$ according to the following optimistic index with respect to the statistics $s_t(i,j)/n_t(i,j)$ for an unranked item $i$ and the item $j$ at the bottom of the leader ranking $\sigmaldrt$ as follows~\cite{gauthier2022unirank}:
\begin{align*}
\doublebars_t(i,j) := 2\times f\left(\frac{1+s_t(i,j)/n_t(i,j)}{2}, n_t(i,j), \tilde{t}_{\sigmaldrt}(t)\right)-1,
\end{align*}
where $\tilde{t}_{\bsigma}(t)$ is the number of rounds where $\bsigma$ has previously been the leader ranking by round $t$ 
and $f$ is the KL-UCB index \cite{cappe2013kullback} of the statistic $\hat{\mu}$:
$f(\hat{\mu},N,t):=\sup\{\mu\in[\hat{\mu},1]:N\times kl(\hat{\mu},\mu)\leq \log(t)+3\log(\log(t))\}$ 
with $kl(\cdot,\cdot)$ the Kullback-Leibler (KL) divergence between two Bernoulli distributions of each mean;
$f(\hat{\mu},N,t)=1$ when $t=0$, $N=0$, or $\hat{\mu}=1$.
Note that $\doublebars_t(i,j)=1$ when $n_t(i,j)=0$.
Then, we can efficiently explore a promising unranked item, which may be superior to the item at the bottom of the current ranking.

\begin{algorithm}[t]
\caption{KL-UCB-BR}
\label{alg:fullbubblerank}
\begin{algorithmic}[1]
\State \textbf{Input:} items $[L]$, initial list $\bsigma_0 \in [L]^K$, $\delta\in(0,1)$
\State  $s_0(i,j) \gets 0$, $n_0(i,j) \gets 0$, $\forall (i,j)\in[L]^2$
\State $\bsigma^{\mathrm{LDR}}_1\gets\bsigma_0$
\For {$t=1,\ldots,T$}
    \State $h \gets t \mod 2$
    \State $\sigmat'\gets\bsigma_t^{\mathrm{LDR}}$
    \State $\sigmat'(K+1) = \argmax_{j\in [L]\setminus\sigmaldrt} \doublebars_{t}(j,\sigmaldrt(K))$
    \State $\sigmat \gets \sigmat'$
    \For{$k =1,\ldots,\lceil(K-h)/2\rceil$}
        \State $i \gets \sigmat(2k-1+h)$, $j \gets \sigmat(2k+h)$
        \If {$s_{t-1}(i,j)\leq 2\sqrt{n_{t-1}(i,j)\log(1/\delta)}$}
            \State Randomly exchange items $\sigmat(2k-1+h)$
            \Statex \hspace{38pt} and $\sigmat(2k+h)$ in list $\sigmat$
        \EndIf
    \EndFor
    \State Display $\sigmat([K])$ and observe clicks $\{c_t(k)\}_{k=1}^K \in \{0,1\}^K$
    \State $s_t \gets s_{t-1}$, $n_t \gets n_{t-1}$
    \For {$k =1,\ldots,\lceil(K-h)/2\rceil$}
        \State $i \gets \sigmat(2k-1+h)$, $j \gets \sigmat(2k+h)$
        \If {|$c_t(2k-1+h)-c_t(2k+h)|=1$}
            \State $s_t(i,j) \gets s_{t}(i,j)+c_t(2k-1+h)-c_t(2k+h)$
            \State $n_t(i,j) \gets n_{t}(i,j)+1$
            \State $s_t(j,i) \gets s_{t}(j,i)-c_t(2k-1+h)+c_t(2k+h)$
            \State $n_t(j,i) \gets n_{t}(j,i)+1$
        \EndIf
    \EndFor
    \For {$k=1,\ldots,K$}
        \State $i\gets\sigmat'(k)$, $j\gets\sigmat'(k+1)$
        \If {$s_{t}(j,i)>2\sqrt{n_{t}(j,i)\log(1/\delta)}$}
            \State Exchange items $\sigmat'(k)$ and $\sigmat'(k+1)$ in list $\sigmat'$
        \EndIf
    \EndFor
    \State $\sigmaldr_{t+1} \gets \sigmat'([K])$
\EndFor
\end{algorithmic}
\end{algorithm}

\section{Experiments}
\subsection{Settings}
We compare KL-UCB-BR to safety-aware and safety-agnostic baseline methods, BubbleRank~\cite{li2020bubblerank}, TopRank~\cite{lattimore2018toprank}, and UniRank~\cite{gauthier2022unirank}, which are click-model-agnostic and do not utilize any item features.
To see the performance gain from the original ranking, we also consider the original ranking as a non-adaptive baseline method; hereafter, it is referred to as OriginalRank.
As we are interested in OLTR problems with unranked items,
we utilize the extension of BubbleRank described by \citet{li2020bubblerank}, in which the item at the bottom of the current ranking is exchanged randomly with a random unranked item, which has not yet been determined to be superior to the bottom-ranked item.

In our experiments, we use the \emph{Yandex} click dataset~\footnote{\url{https://www.kaggle.com/c/yandex-personalized-web-search-challenge}} to simulate user clicks on displayed rankings generated by the algorithms.
This dataset includes over 30 million user sessions, which contain over 20 million unique search queries, extracted from Yandex-search logs.
We basically follow the experimental protocol of the conventional studies~\cite{li2020bubblerank,zoghi2017batchrank}.
To simulate user behavior in the dataset, we estimate the parameters of a click model from the user sessions in the top-100 frequent queries by using PyClick library~\footnote{\url{https://github.com/markovi/PyClick}}.
Throughout experiments, we consider two click models implemented in PyClick, the position-based click model (PBM) and the cascade click model (CM). 
In a single simulated user session,
we generate user clicks on a ranking displayed by an algorithm according to the learned click model and then evaluate the algorithm.

For each query,
we use the most frequent ranking with $10$ items and consider the top-$5$ items in the ranking as the original ranking and the remaining $5$ items as unranked ones.
The goal of the simulation for each query is to rank the top-$5$ most attractive items in descending order of attractiveness among the $10$ ranked/unranked items.
We measure the performance of each algorithm by computing cumulative expected regret defined in~\cref{eq:regret} and safety violation for display rankings.
Safety violation is the number of rounds that an algorithm violates the safety constraint
~\footnote{Our safety constraint is extended from the one proposed by~\citet{li2020bubblerank} to handle unranked items.} 
defined in Definition~\ref{def:safety}.
Violating the safety constraint often leads to user disengagement from applications by displaying users a ranking that is far inferior to the original one.
Notably, KL-UCB-BR and BubbleRank are guaranteed that they do not violate the safety constraint until round $T$ with high probability.
\begin{definition}[Safety Constraint]
\label{def:safety}
Let $\bsigma^* = \{1,2,\ldots,K\}$ be the optimal ranking; $\alpha(1)\geq\alpha(2)\geq\ldots\geq\alpha(L)$; and 
\begin{align*}
V(\bsigma) := |\{(i,j)\in [L]^2: i<j, \bsigma^{-1}(j)<\bsigma^{-1}(i), \bsigma^{-1}(j)\leq K\}|
\end{align*}
be the number of incorrectly-ordered item pairs whose one or both items are in a display ranking $\bsigma$.
Then, the safety constraint for a display ranking $\sigmat$ in round $t$ is
$
V(\sigmat) \leq V(\bsigma_0) + L - K/2,
$
where $\bsigma_0$ is an original ranking.
\end{definition}

\subsection{Results}
In our experiments, we compare KL-UCB-BR with TopRank, UniRank, BubbleRank, and OriginalRank under the click models of PBM and CM. 
As evaluation measures, the cumulative expected regret defined in~\cref{eq:regret} and the safety violation defined in Definition~\ref{def:safety} are computed by taking the average of measurements obtained from $100$ repeated experiments, each with $T=10^5$ rounds.
The shaded regions present standard errors in the measurements.

\begin{figure}[tb]
    \centering
    \includegraphics[width=\linewidth]{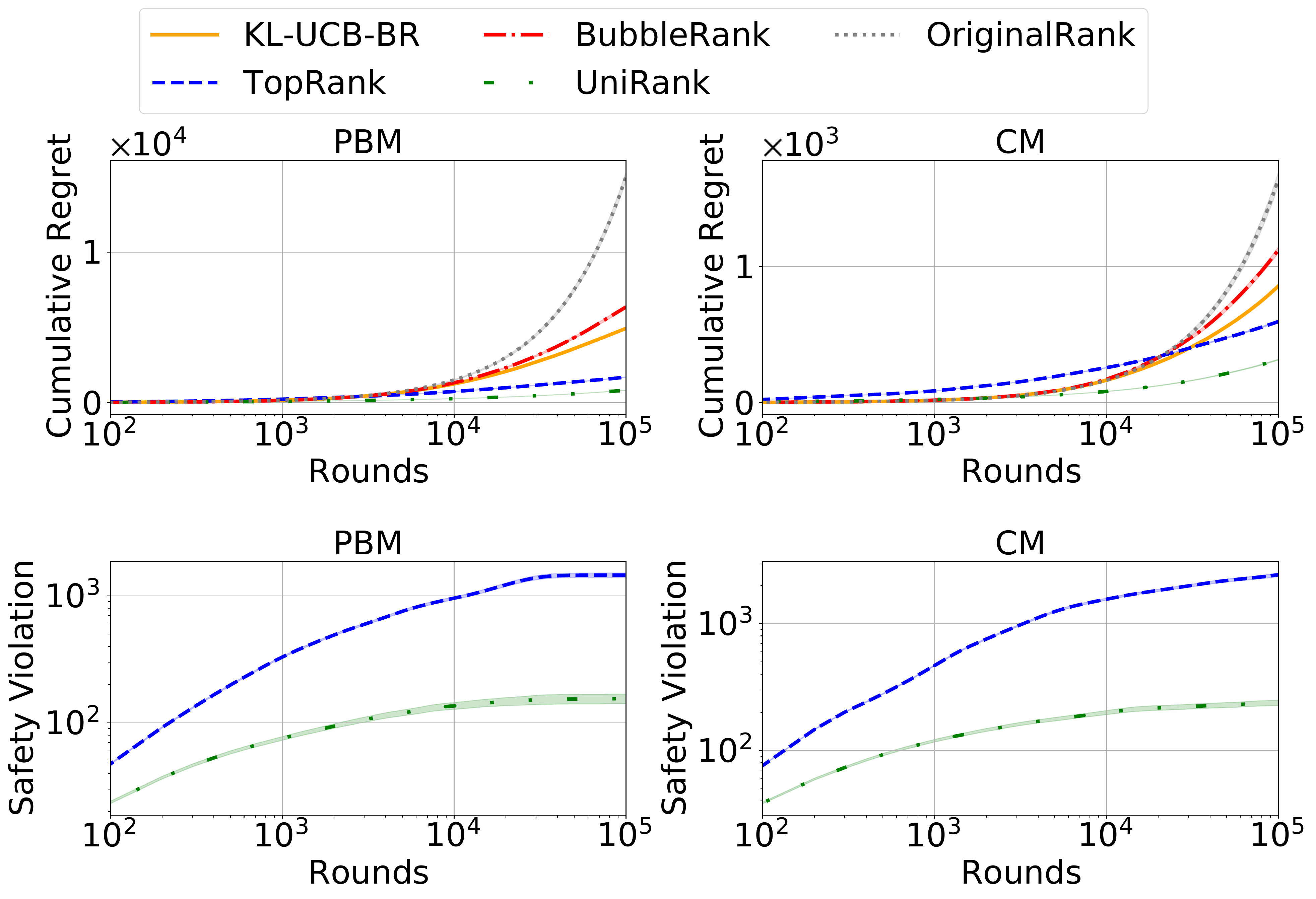}
    \caption{Cumulative expected regret and safety violation with respect to $T=10^5$ rounds for all 100 queries.}
    \label{fig:regret-all}
\end{figure}

\cref{fig:regret-all} shows the results for all 100 queries.
Here, the evaluation measures are computed as averages on the total $10,000$ runs consisting of $100$ experiments for each query.
We can observe that the cumulative expected regret of each adaptive algorithm grows more slowly under PBM and CM than that of OriginalRank (gray) in the two top figures.
This suggests the original rankings have room for improvement for most of the queries.
The top two figures also show that TopRank (blue) and UniRank (green) have lower cumulative expected regret, particularly in the late rounds, under PBM and CM than BubbleRank (red) and KL-UCB-BR (orange), whereas the two bottom figures show that they violate the safety under PBM and CM because of their safety-agnostic exploration.
This implies that these algorithms may lead to user disengagement due to their early-stage unsafe behaviors, and their superiority in terms of long-term regret may not be tangible in practice.
KL-UCB-BR and BubbleRank, on the other hand, have no safety violation by gradually updating an original ranking, thus their plots are invisible.
Among the safe algorithms, KL-UCB-BR in particular has the lowest cumulative expected regrets in the top two figures, suggesting that it can explore unranked items more efficiently than BubbleRank. 

\begin{figure}[tb]
    \centering
    \includegraphics[width=\linewidth]{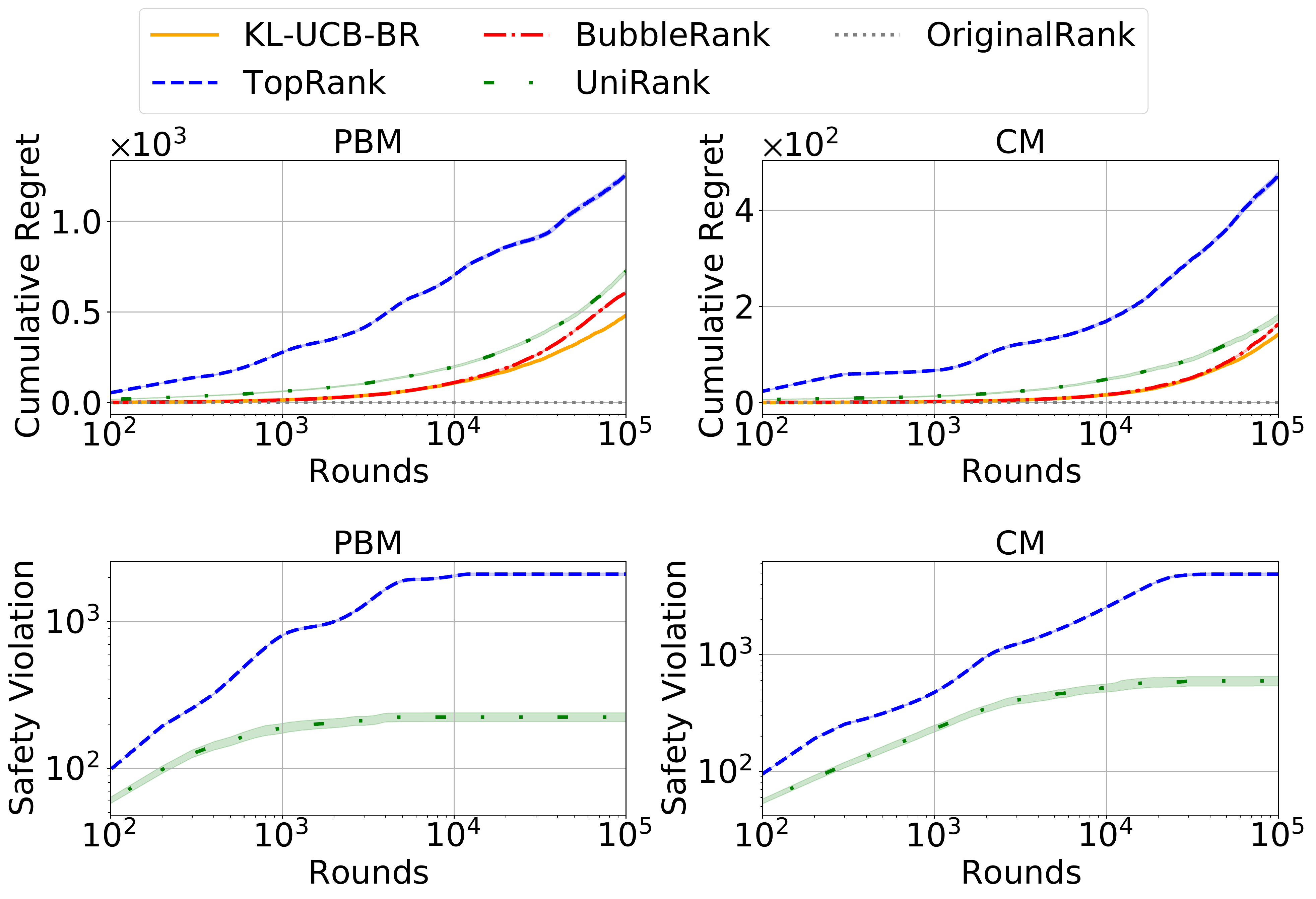}
    \caption{Cumulative expected regret and safety violation with respect to $T=10^5$ rounds for the case when the original ranking is optimal.}
    \label{fig:regret-best}
\end{figure}

\begin{figure}[tb]
    \centering
    \includegraphics[width=\linewidth]{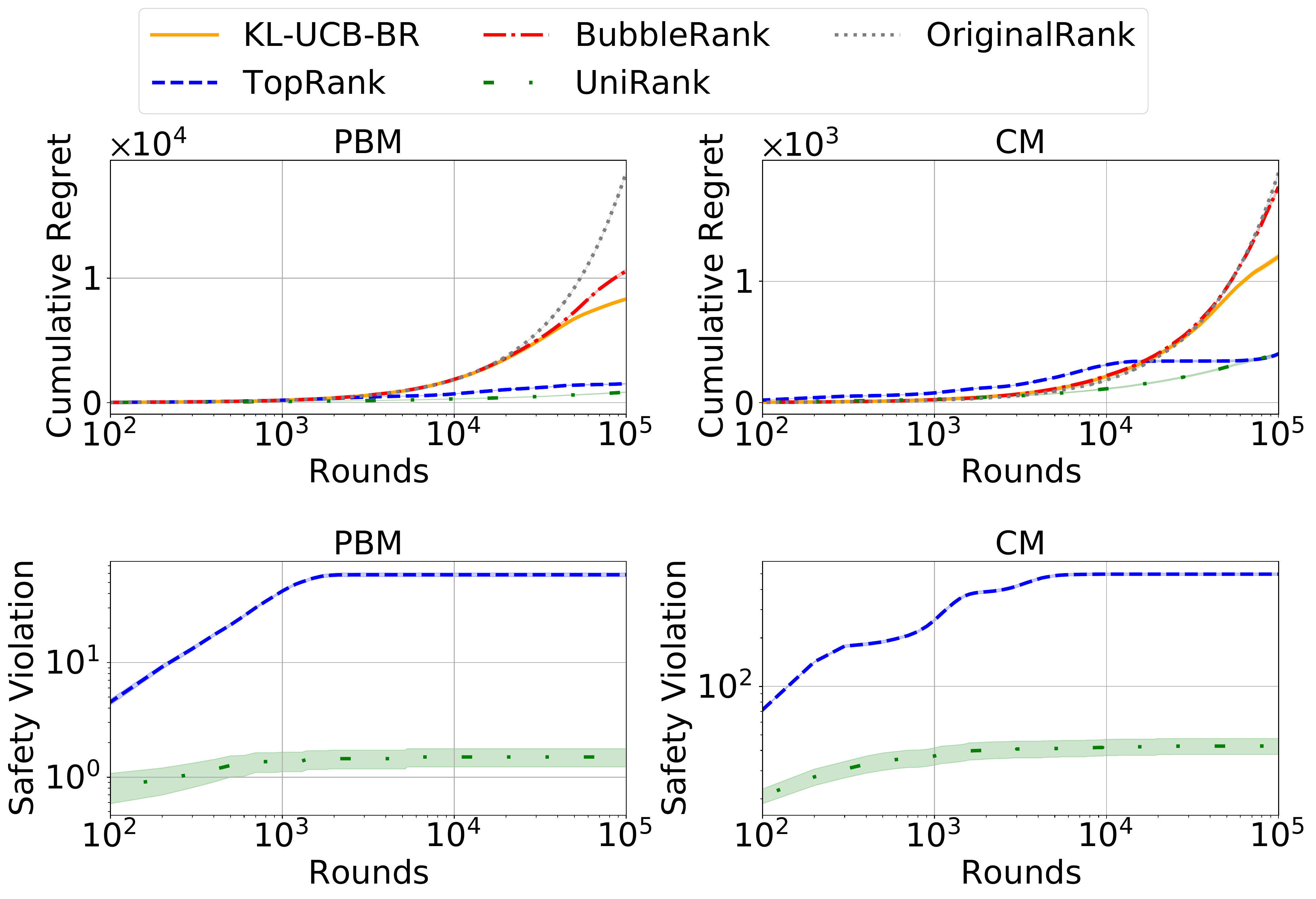}
    \caption{Cumulative expected regret and safety violation with respect to $T=10^5$ rounds for the case when the original ranking does not have the $5$ most attractive items.}
    \label{fig:regret-worst}
\end{figure}

Second, we report the results for each characteristic query.
\cref{fig:regret-best} shows the results for the query whose original ranking is optimal; namely, it has the $5$ most attractive items in descending order of their attractiveness.
Note that, in this situation, regrets still occur when an algorithm explores ranked and unranked items.
As a result, OriginalRank (gray) has no regret in the entire rounds in the two top figures.
KL-UCB-BR (orange) and BubbleRank (red) have lower cumulative expected regrets under PBM and CM than TopRank (blue) and UniRank (green) because they update their display rankings under the safety constraint with respect to the original ranking (see the bottom two figures) and thereby avoiding destructive regret.
Among safe algorithms, KL-UCB-BR demonstrates lower cumulative expected regret in the late rounds than BubbleRank since it can cut losses stemming from unranked items with low attractiveness.
\cref{fig:regret-worst} shows the results of the query whose original ranking does not include the $5$ most attractive items.
The top two figures show that KL-UCB-BR and BubbleRank have higher cumulative expected regrets under PBM and CM than TopRank and UniRank.
This is because the safety-aware methods avoid updating the original ranking aggressively and thus suffer from regret due to the less-than-optimal original ranking.
Still, we can take advantage of the safety-aware methods in practical applications as the cumulative expected regrets are improved from OriginalRank to some extent.
In particular, KL-UCB-BR outperforms BubbleRank in terms of cumulative expected regret; the performance gain of KL-UCB-BR is emphasized in the late rounds.
It is also remarkable that, under CM, KL-UCB-BR shows considerable improvement from OriginalRank, while BubbleRank fails to reduce regret from that of OriginalRank.
These results represent that the KL-UCB-based exploration can find some of the $5$ most attractive items from unranked items while random exploration can not find them because regrets do not occur by ordering items incorrectly in display rankings under CM.


\section{Conclusion}
We propose KL-UCB-BR, which is a safe OLTR algorithm that can explore both ranked and unranked items without violating its safety constraint. 
KL-UCB-BR can efficiently explore unranked items by using the KL-UCB index 
in contrast to the original BubbleRank with random exploration~\cite{li2020bubblerank}. 
We empirically demonstrate that the KL-UCB-BR outperforms BubbleRank without any safety violation in various scenarios.

\onecolumn
\begin{multicols}{2}
    \bibliographystyle{ACM-Reference-Format}
    \bibliography{safeoltr.bib}
\end{multicols}

\end{document}